\documentclass[twocolumn,showpacs,nopreprintnumbers,amsmath,amssymb,prl]{revtex4}
\usepackage{color}
\usepackage{graphicx}
\usepackage{dcolumn}
\usepackage{bm}

\begin{document}

\hyphenation{acce-le-ra-tion tomo-graphy prio-ri maxi-mum}
\newcommand{\etal}{\textit{et al. }}
\newcommand{\mat}[1]{{\mathbi{#1}}}
\newcommand{\vect}[1]{\mathbf{#1}} 
\newcommand{\di}{d} 
\newcommand{\e}{e} 
\newcommand{\I}{i} 

\newcommand{\thisarticle}{this article}

\title{Molecular Orbital Tomography using Short Laser Pulses}

\author{Elmar V. \surname{van der Zwan}}
 \email{zwan@physik.uni-kassel.de}

\author{Ciprian C. Chiril\u{a}}
\author{Manfred Lein}
\affiliation{
Institute for Physics and Center for Interdisciplinary Nanostructure Science and Technology, University of Kassel, Heinrich-Plett-Stra{\ss}e 40, 34132 Kassel, Germany
}%

\date{\today}

\begin{abstract}
Recently, a method to image molecular electronic wave functions using high harmonic generation (HHG) was introduced by Itatani \etal [Nature {\textbf{432}}, 876 (2004)]. We show that, while the tomographic reconstruction of general orbitals with arbitrary symmetry cannot be performed with long laser pulses, this becomes possible when extremely short pulses are used. An alternative reconstruction equation based on momentum matrix elements, rather than on dipole matrix elements, is proposed. We present simulations of the procedure for 2D model systems based on numerical solutions of the time-dependent Schr\"{o}dinger equation, and present results from further post-processing of the reconstructed orbitals. 
\end{abstract}


\maketitle

High-order harmonic generation (HHG) stands for the emission of high-frequency radiation from a system 
driven by a strong laser field.
The three-step picture \cite{Corkum93} 
explains HHG for atoms and molecules by a sequence
of field ionization, acceleration of the free electron in the laser field 
and recollision with the core. HHG has received increasing attention in the last years, since it can be used both as a source of coherent radiation, e.g., for creating attosecond pulses \cite{Drescher01,Paul01,Mairesse03,Tzallas03}, and as a diagnostic tool to determine molecular properties such as the internuclear distance \cite{Lein02,Kanai05,Baker06,Lein07}. Recently, HHG has been employed to image electronic orbitals \cite{Itatani04}, a scheme known as molecular orbital tomography. There, it was shown that the electronic orbital of N$_2$ (including the sign of the wave function) can be reconstructed. The method is in principle also suited for observing femtosecond electron dynamics in chemical reactions. The potential of HHG as a femtosecond probe has been proven in experiments on vibrating SF$_6$ molecules \cite{Wagner06}.

The main idea behind molecular orbital tomography is that the returning electron wave packet 
in the three-step picture can be approximately regarded as a plane wave. The transition matrix element describing the recombination is then
a Fourier transform of the bound state. 
The spectra from many different orientations of the molecules are combined to reconstruct a 2D projection of the bound-state orbital on the plane orthogonal to the pulse propagation direction. The necessary information about the continuum wave packet is approximately obtained by measuring the spectrum of a 
reference atom with the same ionization potential as the molecular system. 

In a multi-electron system, corrections beyond the single-active electron model should be taken into account \cite{Gordon06-2}. As a consequence, a modified Dyson orbital and not the highest occupied molecular orbital is reconstructed \cite{Santra06, Patchkovskii06,Patchkovskii07}. Recently it has been argued that the structure of the continuum wave packet does not vary between different rare-gas atoms \cite{Levesque07}. Also it was shown that the orientation dependence of HHG from the more complex molecules acetylene and allene can still be understood in a single-active-electron approximation \cite{Torres07}. Both observations support the idea of extending the tomographic technique to molecules other than N$_2$. 

In the current work, we present an analysis of the method with important conclusions for the application of molecular orbital tomography to more complex systems. In contrast to the original work \cite{Itatani04}, we use momentum rather than dipole matrix elements \cite{vanderZwan07} (i.e., \textit{velocity form} instead of \textit{length form}), as the velocity form is better suited to quantify the recombination step in HHG \cite{Chirila07}. We show that the reconstruction of an arbitrary orbital (without symmetry) is impossible within the original scheme based on many-cycle pulses. Instead, one must ensure that the electron wave packets return to the core from only one side. We propose to achieve this by using extremely short laser pulses. It is important to note that uni-directional recollision is a requirement additional to the need of molecular head-versus-tail orientation. Furthermore we show that using  \textit{a priori} assumptions about the type of errors present in the method, to some extent those errors can be iteratively removed from the results. 

Let us consider a molecular system in a laser field polarized in the $x$-direction and propagating in the $z$-direction. 
The intensity of the harmonic radiation with frequency $\omega$ 
is proportional to
\begin{equation}
I(\omega) = \omega^2(|\vect{P}(\omega)|^2+|\vect{P}(-\omega)|^2).
\end{equation}
The phases of the radiation are given by 
$\arg[\vect{P}(\omega)]$. Atomic units are used throughout \thisarticle . Here, $\vec{P}(\omega)$ is the Fourier transformed dipole velocity
\begin{subequations}
\begin{align}
\vect{P}(\omega) &= \int \vect{p}(t) \e^{\I \omega t} \di t \label{PwFTEquation},\\
\vect{p}(t) &= -\langle \psi(x,y,z,t)|\hat{\vect{p}}|\psi(x,y,z,t)\rangle,
\end{align}
\end{subequations}
with $\hat{\vect{p}}=(-\I \partial_x,-\I \partial_y)$ denoting the electron momentum operator in the $xy$-plane. This means that we restrict ourselves to harmonics emitted along the z-axis. In the single-active-electron model, the time-dependent wave function $\psi$ is split into two parts as $\psi(x,y,z,t)=\psi_0(x,y,z,t)+\psi_{\mathrm{c}}(x,y,z,t)$, where $\psi_0(x,y,z,t)$ is the initial bound-state wave function and $\psi_{\mathrm{c}}(x,y,z,t)$ is the continuum wave packet. The time-dependence of the initial state is given by \mbox{$\psi_0(x,y,z,t)=\psi_0(x,y,z) \exp{(\I I_{\mathrm{p}} t)}$},
where $I_{\mathrm{p}}$ is the ionization potential of the initial state. 

To show the need for unidirectional recollision, we first give a simple argument based on the laser-field-free recombination of an electron wave packet. Introducing the plane-wave approximation, a continuum wave packet can be written as
\begin{equation}
\psi_{\mathrm{c}}(x,y,z,t) = \int_{-\infty}^\infty  a(k) \e^{\I k x} \e^{-\I \frac{k^2}{2} t} \frac{\di k}{2\pi},
\label{psictEquation}
\end{equation}
where the $a(k)$ are complex amplitudes. Neglecting depletion of the initial state, the momentum expectation value $\vect{p}(t)$ is given by
\begin{equation}
\vect{p}(t) \simeq  -\langle \psi_0(x,y,z,t)|\hat{\vect{p}}|\psi_{\mathrm{c}}(x,y,z,t)\rangle + \mathrm{c.c.}
\label{eq: p(t)}
\end{equation}
In the frequency domain we obtain for $\omega>0$
\begin{equation}
\begin{split}
\vect{P}(\omega) &= -\int_{-\infty}^\infty a(k) \iint  \psi_0^{\mbox{{\footnotesize 2D}}}(x,y) \hat{\vect{p}}\; \e^{\I k x} \di x \di y \\
&\quad  \times \delta(\omega-I_{\mathrm{p}}-\frac{k^2}{2}) \di k,
\end{split}
\label{PwFirstEquation}
\end{equation}
with \mbox{$\psi_0^{\mbox{{\footnotesize 2D}}}(x,y)=\int \psi_{0}(x,y,z)\di z$}. The $y$-component of $\vect{P}$ is identically zero. This means that the velocity form, despite yielding good results for tomography, is not suitable to predict the harmonic polarization correctly. For \mbox{$\omega > I_{\mathrm{p}}$}, we obtain
\begin{subequations}
\begin{align}
I(\omega)&= 2\omega^2 \left|P_{x}(\omega)\right|^2,\\
P_{x}(\omega) &= -a(k(\omega))\;p^*(\omega)+a(-k(\omega))\;p(\omega),\label{eq: Pandp}\\
p(\omega) &= \iint \psi_0^{\mbox{{\footnotesize 2D}}}(x,y) \e^{-\I k(\omega) x} \di x \di y,
\end{align}
\end{subequations}
where we have defined the wave number $k(\omega)=\sqrt{2(\omega-I_{\mathrm{p}})}$. To obtain $p(\omega)$ from known $P_{x}$ and $a(k)$, we need to combine the two terms on the right-hand side of Eq.~\eqref{eq: Pandp} into one term, such that $p(\omega)$ is factored out. There are two cases in which this is possible: (i) the orbital is \textit{\mbox{(un-)gerade}} or (ii) $a(k)=0$ for $k>0$ (or $k<0$), i.e., the returning wave packets approach the nuclei always from the same side. In the first case, we have $p^*(\omega)=\pm p(\omega)$ and the experimentally determined amplitudes will be a linear combination of $a(k)$ and $a(-k)$. In this paper we focus on the second possibility. Assuming that recollision occurs only from $x>0$, the Fourier transformed dipole velocity is given by
\begin{equation}
P_x(\omega)= a(-k(\omega)) p(\omega).
\label{eq: PShortPulse}
\end{equation}
As a modification to the above equation we use the modified wave number $k(\omega) = \sqrt{2\omega}$, as in \cite{Itatani04}. The modification is related to the fact that high harmonics seem to be better described when the ionization potential is left out of the relation \cite{Lein02}. This is also supported by numerical tests that we have performed, although for SFA-based simulations it was recently reported that the original relation should probably be used \cite{Le07, Levesque07}. Physically, the argument is that when describing the returning wave packet as a plane wave, we should take into account that
at the moment of recombination its wave number is modified due to the kinetic energy being increased by an amount of the order of $I_{\mathrm{p}}$. This amounts to the WKB-approximation for the returning wave packet \cite{Zhao07}. Although it is becoming experimentally possible to measure both the amplitude and the phase of the harmonics \cite{Mairesse03,Sekikawa03, Varju05, Mairesse05, Kanai07}, in the original work \cite{Itatani04} only the intensities were measured. The phase information was added from \textit{a priori} considerations. 

There are two unknowns in our set of equations, namely $a(k(\omega))$ and $\psi_{0}^{\mbox{{\footnotesize 2D}}}(x,y)$. One of the main ideas of the procedure is to solve this problem by comparison with an atomic reference system for which $\psi_0^{\mathrm{(a)}}(x,y)$ is known and $a^{\mathrm{(a)}}(k)$ is very similar to $a(k)$. (With the superscript `(a)' we denote reference-system quantities.) This is approximately the case if the reference system and the molecule of interest have the same ionization potential \cite{Itatani04}. 

In the experimental implementation \cite{Itatani04}, the molecules are aligned along directions within the $xy$-plane, with an angle $\theta$ between the molecular axis and the electric field. Harmonic generation is then considered for all orientations $\theta$ and determined by the projections of the bound state \mbox{$\psi_{0,\theta}^{\mbox{{\footnotesize 2D}}}(x,y)=\psi_{0}^{\mbox{{\footnotesize 2D}}}(x \cos\theta+y \sin\theta,-x \sin\theta+y\cos\theta)$}. The general Fourier transform
\begin{equation}
g(k_1,k_2) = \iint \psi_{0}^{\mbox{{\footnotesize 2D}}}(x,y) \e^{-\I (k_1 x +k_2 y)} \di x \di y
\end{equation}
can be inverted to get
\begin{equation}
\psi_{0}^{\mbox{{\footnotesize 2D}}}(x,y)  = \frac{1}{(2\pi)^2}\iint g(k_1,k_2) \e^{\I (k_1 x +k_2 y)} dk_1 dk_2.
\label{psi0k1k2}
\end{equation}
We substitute $k_1=k\cos\theta$ and $k_2=-k\sin\theta$ and change the integration variables accordingly, and use $p_{\theta}(\omega)=g(k(\omega) \cos\theta,-k(\omega) \sin\theta)$ to arrive at
\begin{equation}
\psi_{0}^{\mbox{{\footnotesize 2D}}}(x,y)= \iint p_{\theta}(\omega) \e^{\I k(\omega)(x \cos \theta -y \sin\theta)}\frac{d\omega d\theta}{(2\pi)^2}.
\label{eq: psi02D}
\end{equation}
Using Eq.~\eqref{eq: psi02D}, one can reconstruct $\psi_{0}^{\mbox{{\footnotesize 2D}}}$ from $ p_{\theta}(\omega)$. 

A more rigorous analysis of the tomography scheme is based on the Lewenstein model \cite{Lewenstein94}. Using the velocity form \cite{Chirila07}, and a laser pulse that is turned on after \mbox{$t=0$}, and turned off before \mbox{$t=T_{\mathrm{p}}$}, the dipole velocity \mbox{$P_{x,\theta}(\omega>0)$} reads
\begin{equation}
\begin{split}
P_{x,\theta}&(\omega) = -\I \int_{0}^{T_{\mathrm{p}}} dt \int_0^{t} d t' \\ &\times v_\theta^{*}(k_{s}(t,t')+A(t))  \e^{-\I S(t,t') + \I \omega t}\\ &\times 
   d_{\mathrm{ion},\theta}(k_{s}(t,t')+A(t'),t') \biggl[\frac{2\pi}{\epsilon + \I(t-t')}\biggr]^{3/2},
\label{eq: SFAacc}
\end{split}
\end{equation}
where we have neglected the `c.c.'-term in Eq.~\eqref{eq: p(t)}. The saddle-point momentum is given by \mbox{$k_{\mathrm{s}}(t,t')=-\int_{t'}^{t} A(t'') dt''/(t-t')$}.  Here \mbox{$A(t)=-\int_{-\infty}^{t} dt' E(t')$} and $E(t)$ is the electric field of the laser pulse. The matrix elements describing the ionization and recombination, and the semi-classical action are given by
\begin{subequations}
\begin{align}
&d_{\mathrm{ion},\theta}(k,t) =\iint \psi_{0,\theta}(x,y)\; E(t) x\; \e^{-\I k x}  \frac{\di x \di y}{(2\pi)^{3/2}},\\
&v_\theta(k)=\iint  \e^{-\I k x} \; \I \partial_x \; \psi_{0,\theta}(x,y) \frac{\di x \di y}{(2\pi)^{3/2}},\\
&S(t,t') =  \int_{t'}^{t} dt'' \; \frac{[k_{\mathrm{s}}(t,t')+A(t'')]^2}{2}
 + I_{\rm p}(t-t').
\end{align}
\end{subequations}
For the tomographic procedure, it is required that in Eq.~\eqref{eq: SFAacc} the recombination matrix element be factored out. To this end, we perform the integration over the recombination time $t$ using the saddle-point approximation. The saddle-point condition 
\begin{equation}
(k_{\mathrm{s}}(t_{\mathrm{s}},t')+A(t_{\mathrm{s}}))^2/2=\omega-I_{\mathrm{p}}
\label{eq: returncondition}
\end{equation}
has the form of an energy-conservation law. We assume that the electron wave packets approach the core always from the positive side, i.e., with negative momentum along the polarization direction. With Eq.~\eqref{eq: returncondition}, the return momentum is given by $-k(\omega)$, where $k(\omega)=\sqrt{2(\omega-I_\mathrm{p})}$. We obtain
\begin{equation}
\begin{split}
P&_{x,\theta}(\omega) = \frac{-\I k(\omega)}{(2\pi)^{3/2}} \; p_\theta(\omega)  \int_0^{T_{\mathrm{p}}} dt' \sum_{t_{\mathrm{s}}>t'} \e^{-\I S(t_{\mathrm{s}},t')}\\ &\times  \e^{\I \omega t_{\mathrm{s}}} \sqrt{2\pi \I/ \left[\frac{2(\omega-I_{\mathrm{p}})}{t_{\mathrm{s}}-t'}-E(t_{\mathrm{s}})k(\omega)\right]}
   \\ &\times d_{\mathrm{ion},\theta}(k_{\mathrm{s}}(t_{\mathrm{s}},t')+A(t'),t')  \biggl[\frac{2\pi}{\epsilon + \I(t_{\mathrm{s}}-t')}\biggr]^{3/2} .
\label{eq: finalPx}
\end{split}
\end{equation}
 In this case molecular orbital tomography is possible and comparison with Eq.~\eqref{eq: PShortPulse} shows that $a(-k(\omega))$ becomes a sum over all classical trajectories leading to emission of a photon with energy $\omega$. Within the SFA, $a(k)$ is identical for length- and velocity-form reconstruction. Note that the structure of Eq.~\eqref{eq: finalPx}, with $p_\theta(\omega)$ as a prefactor of the integral, would not emerge if both positive and negative momenta were present (long-pulse case).   

\begin{figure}[htbp]
  \centering
  \includegraphics[width=0.8\columnwidth]{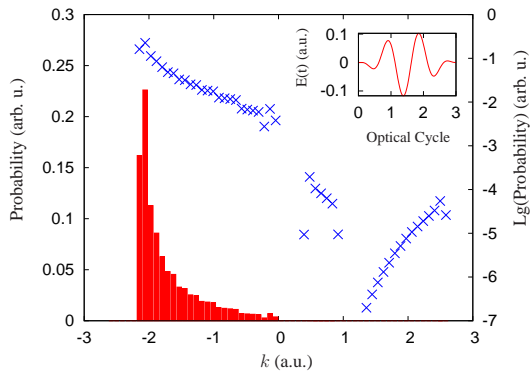}
  \caption{(color online). Semi-classical distribution of return momenta on a linear scale (bars) and on a logarithmic scale (crosses) for the laser intensity $5\times 10^{14}$ W/cm$^2$ and wavelength $780$ nm. Electrons with an energy of 3.17 times the ponderomotive potential have $k\simeq \pm 2.57$ a.u. In the inset the time-dependent electric field of the laser pulse is shown.}
  \label{fig: velspread}
\end{figure}

To find a pulse for which the associated continuum wave packets approach the core from only one side, we calculate semi-classically the probability that an electron returns to the core with momentum $k$, as introduced in \cite{vanderZwan07-3}. The return probability is the classical equivalent of $|a(k)|^2$. After tunneling, the electrons follow classical trajectories. We sample the ionization time and record the trajectories that return within a certain velocity interval. Each trajectory is weighted based on the tunneling probability and wave-packet spreading from the Lewenstein model \cite{Lewenstein94}. A 3-cycle $\sin^2$-pulse with a carrier-envelope phase of $1.25\pi$, for which the results are shown in Fig.~\ref{fig: velspread}, is found to be a good candidate. 
To ensure that effectively only a half-cycle of the pulse contributes to the spectrum, alternative approaches are conceivable. An example is using the ellipticity of the pulse, similar to the polarization gating method in attosecond pulse production \cite{Corkum94,Sola06}.

\begin{figure}[htbp]
\centerline{
\includegraphics[width=0.9\columnwidth,angle=-90]{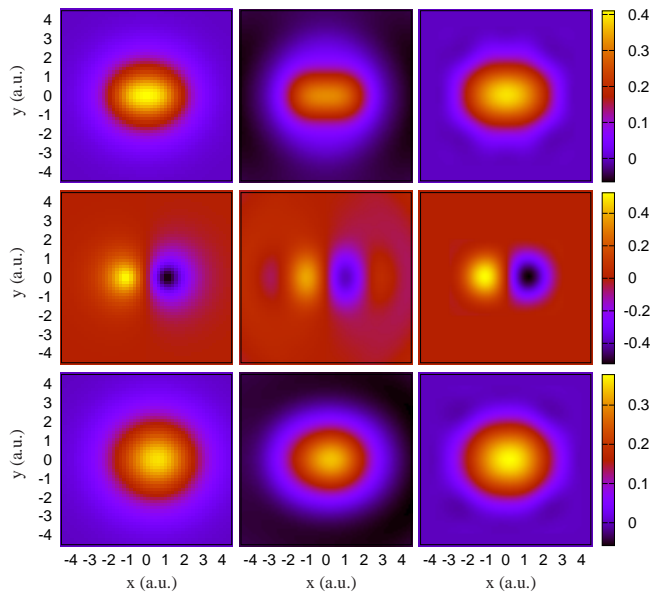}}
\caption{(color online). Simulations of orbital tomography. From top to bottom: the ground state of 2D H$_2^+$ with \mbox{$R=2.0$ a.u.}, its first excited state and the ground state of 2D (H-He)$^{2+}$ with \mbox{$R=2.2$ a.u.} From left to right: the exact bound-state orbital, the reconstructed orbital and the optimized orbital after post-processing. }
\label{fig: results}
\end{figure}

For the simulations of orbital tomography, we study three cases: a symmetric orbital (ground state of 2D H$_2^+$ with the equilibrium internuclear distance of $R=2.0$ a.u.\  as in \cite{Lein02}), an antisymmetric orbital (the first excited state of the same system) and a prototype asymmetric orbital, for which we use the ground state of 2D (H-He)$^{2+}$ with fixed internuclear distance \mbox{$R=2.2$} a.u. We use a laser intensity of \mbox{$5\times 10^{14}$ W/cm$^2$} and wavelength \mbox{$780$ nm}. We adjusted the softcore parameters $a^2$ of the softcore potentials such that $I_{\mathrm{p}}$ equals 30.2 eV in all cases. For the asymmetric state the potential takes the form
\begin{equation}
V(\vect{r})=-\frac{1}{\sqrt{(\vect{r}+\tfrac{\vect{R}}{2})^2+a^2}}-\frac{2}{\sqrt{(\vect{r}-\tfrac{\vect{R}}{2})^2+a^2}}.
\end{equation}
Although the results for the antisymmetric orbital could be improved by using harmonic phases as calculated from the two-center interference model \cite{Lein02}, we choose to present here the results using numerically calculated harmonic phases. First the harmonic spectra and phases are calculated by solving the TDSE for (i) many different orientations of the orbital and (ii) the reference system, 2D He$^+$ for the H$_2^+$ states and 2D Li$^{2+}$ for (H-He)$^{2+}$, also with softcore parameters adjusted such that the ionization potential is 30.2 eV.  Then for velocity-form reconstruction $a_\theta(k)$ is calculated according to 
\begin{equation}
a_\theta(-k(\omega))= \frac{P_{x,\theta}^{\mathrm{(a)}}(\omega) }{ p_\theta^{\mathrm{(a)}}(\omega)}\sqrt{\frac{P_{\mathrm{I}}(\theta)}{P_{\mathrm{I}}^{\mathrm{(a)}}}},
\end{equation}
where $P_{\mathrm{I}}^{\mathrm{(a)}}$ is the ionization yield of the reference system and $P_{\mathrm{I}}(\theta)$ is the ionization yield for different orientations of the molecule. The Fourier transforms $p_\theta(\omega)$ are obtained from Eq.~\eqref{eq: PShortPulse} and the reconstruction is performed using Eq.~\eqref{eq: psi02D}. The mirror-symmetries present in the orbitals have been used to simplify the reconstructions. After the tomographic reconstruction, post-processing procedures can be applied to improve the result. We propose an error-reduction scheme based on the hybrid-input-output algorithm for phase retrieval of Fourier transforms introduced by Fienup \cite{Fienup78}. The reconstructed image is transformed between Fourier and real space iteratively, and in both spaces filters are applied that follow from known properties of the orbital. One assumption is that the systematical error present in the reconstructed image has limited $\theta$- and $k$-dependence. Other important filters include the limited area around the origin where the wave function is allowed to be non-zero (the support area) and the realness of the orbital. We will report more extensively about the technique elsewhere. The results are shown in Fig.~\ref{fig: results}. In the cases studied here, the tomographic reconstruction already yields orbitals that resemble well the exact orbitals. The error-reduction scheme improves further on these results. 

In summary, we have carried out an analysis of molecular orbital tomography. Using velocity-form reconstruction and uni-directional wave packets obtained with short pulses, orbitals of arbitrary symmetry can be reconstructed. A further improvement is possible by post-processing the reconstructed image. Molecular orbital tomography represents an interesting and promising technique that bears great potential for observing femtosecond dynamics of electronic orbitals.

We acknowledge discussions within the NSERC SRO network ``Controlled electron re-scattering: femtosecond, sub-angstrom imaging of single molecules''. This work was partially supported by the Deutsche Forschungsgemeinschaft.

\bibliography{/home/theo/zwan/articles/references}
\end{document}